\documentclass[apj]{emulateapj}
\usepackage{longtable}

\newcommand{\kms}         {km~s$^{-1}$}

\shorttitle{High-Resolution Spectroscopy of Leo IV}
\shortauthors{Simon et al.}  
\slugcomment{ApJ, in press}

\begin{document}

\title{High-Resolution Spectroscopy of Extremely Metal-Poor Stars in
  the Least Evolved Galaxies: Leo IV\altaffilmark{*}}

\author{Joshua D. Simon\altaffilmark{1}, Anna Frebel\altaffilmark{2},
  Andrew McWilliam\altaffilmark{1}, Evan
  N. Kirby\altaffilmark{3,4}, and Ian B. Thompson\altaffilmark{1}}

\altaffiltext{*}{This paper includes data gathered with the 6.5 meter
  Magellan Telescopes located at Las Campanas Observatory, Chile.}

\altaffiltext{1}{Observatories of the Carnegie Institution of
  Washington, 813 Santa Barbara St., Pasadena, CA 91101;
  jsimon@ociw.edu, andy@ociw.edu, ian@ociw.edu}

\altaffiltext{2}{Harvard-Smithsonian Center for
  Astrophysics, Cambridge, MA 02138; afrebel@cfa.harvard.edu}

\altaffiltext{3}{Department of Astronomy, 
       California Institute of Technology, 1200 E. California Blvd.,
       MS 249-17, Pasadena, CA  91125; enk@astro.caltech.edu }

\altaffiltext{4}{Hubble Fellow}

\begin{abstract}
We present high-resolution Magellan/MIKE spectroscopy of the brightest
star in the ultra-faint dwarf galaxy Leo~IV.  We measure an iron
abundance of $\rm{[Fe/H]} = -3.2$, adding to the rapidly growing
sample of extremely metal-poor stars being identified in Milky Way
satellite galaxies.  The star is enhanced in the $\alpha$ elements Mg,
Ca, and Ti by $\sim0.3$~dex, very similar to the typical Milky Way
halo abundance pattern.  All of the light and iron-peak elements
follow the trends established by extremely metal-poor halo stars, but
the neutron-capture elements Ba and Sr are significantly
underabundant.  These results are quite similar to those found for
stars in the ultra-faint dwarfs Ursa Major II, Coma Berenices,
Bo{\"o}tes~I, and Hercules, suggesting that the chemical evolution of
the lowest luminosity galaxies may be universal.  The abundance
pattern we observe is consistent with predictions for nucleosynthesis
from a Population III supernova explosion.  The extremely low
metallicity of this star also supports the idea that a significant
fraction ($\gtrsim10$\%) of the stars in the faintest dwarfs have
metallicities below $\rm{[Fe/H]} = -3.0$.
\end{abstract}


\section{INTRODUCTION}

The chemical abundance patterns of the most metal-poor stars provide a
unique fossil record of star formation, chemical evolution, and
supernova nucleosynthesis in the early universe.  Until recently, such
studies were limited to the stellar halo of the Milky Way because
nearby dwarf galaxies appeared to lack sufficiently metal-poor stars
\citep{helmi06}.  Just over a year ago, the first extremely metal-poor
(EMP) stars with $\rm{[Fe/H]} \le -3.0$ were discovered in several of
the Milky Way's lowest luminosity companions \citep{kirby08}.  Since
then, the number of known EMP stars in nearby dwarf galaxies has been
expanding rapidly \citep*{frebel09,ch09,aoki09,fks09,norris09}.

Because the ultra-faint dwarfs host such incredibly tiny stellar
populations ($L \lesssim 10^{4}$~L$_{\odot}$), they represent
particularly attractive targets for chemical abundance studies.  These
galaxies should have hosted only a few supernovae (SNe), and the
individual chemical signatures of those explosions may be revealed in
their oldest stars \citep[e.g.,][]{koch08}.  Moreover, they were
likely some of the first objects to collapse in the early universe
\citep{br09}, and the lack of star formation at later times means that
evidence of the nucleosynthetic processes operating at high redshift
should be preserved.

In a previous paper we presented high-resolution spectra of six stars
in the ultra-faint dwarfs Ursa Major~II (UMa~II) and Coma Berenices
(ComBer), showing that both galaxies have very low metallicities,
substantial iron abundance spreads, and overall abundance patterns
similar to that of the Milky Way halo \citep{frebel09}.  Here we
report spectroscopy of the brightest star (and the only one accessible
to current telescopes at high spectral resolution) in the slightly
more luminous galaxy Leo~IV.  In \S~\ref{observations} we describe
Leo~IV and our observations.  We present our abundance analysis in \S
\ref{results}, and then discuss the implications of our results for
the chemical evolution of the faintest galaxies in
\S~\ref{discussion}.  In \S~\ref{conclusions} we briefly summarize our
findings and conclude.

\section{OBSERVATIONS AND DATA REDUCTION}
\label{observations}

\subsection{Properties of Leo~IV}

Leo~IV was discovered as an overdensity of resolved stars in the fifth
data release of the Sloan Digital Sky Survey \citep{sdss_dr5} by
\citet{belokurov07}.  Medium-resolution spectroscopy by
\citet[][hereafter SG07]{sg07} demonstrated that Leo~IV has stellar
kinematics and metallicities that are characteristic of dwarf
galaxies, but as the most poorly-studied object in the SG07 sample its
overall properties were not well constrained.  Followup analysis of
the SG07 spectra by \citet{kirby08} revealed that Leo~IV has the
lowest mean metallicity of any galaxy known, at $\rm{[Fe/H]} = -2.58
\pm 0.08$, with a very large internal metallicity spread of 0.75~dex.
Subsequently, photometric studies by \citet*{martin08},
\citet{sand09}, \citet{moretti09}, and \citet{dejong09} refined the
size ($128 \pm 26$~pc), absolute magnitude ($M_{V} = -5.7 \pm 0.3$),
and distance ($154 \pm 5$~kpc) of the galaxy.  SG07 identified a
single bright red giant star, SDSSJ113255.99--003027.8 (hereafter
referred to as Leo~IV-S1), in Leo~IV at $V=19.2$, with the next
brightest confirmed member nearly a magnitude fainter.

\subsection{Observations}

We observed Leo~IV-S1 on 2009 February 18--20 with the Magellan
Inamori Kyocera Echelle (MIKE) spectrograph \citep{mike} on the Clay
Telescope.  The observations were made with a 1\arcsec\ slit,
producing a spectral resolution of $R = 28,000$ on the blue side
($\lambda < 5000$~\AA) and a resolution of $R = 22,000$ on the red
side ($\lambda > 5000$~\AA).  The CCD was binned $3 \times 3$ to
reduce read noise for such a faint target, yielding a final dispersion
of $\sim0.07$~\AA~pixel$^{-1}$ in the blue and
$\sim0.12$~\AA~pixel$^{-1}$ in the red (i.e., sampling slightly better
than the Nyquist rate).  A temporary, lower efficiency detector was
used because of the failure of the MIKE blue CCD in 2008 November. The
total integration time was 8.67 hours (individual exposures were
either 40 or 55 minutes) under mostly excellent observing conditions.

\subsection{Data Reduction}

The data were reduced using the latest version of the MIKE pipeline
introduced by \citet{kelson03}.  Frames from each night were reduced
together, and then the spectra from the individual nights were coadded
at the end.  The final spectrum was normalized in IRAF\footnote{IRAF
  is distributed by the National Optical Astronomy Observatory, which
  is operated by the Association of Universities for Research in
  Astronomy under cooperative agreement with the National Science
  Foundation.} and each order was analyzed separately.  Because of the
target star's faintness, the signal-to-noise ratio (S/N) achieved is
modest: $\rm{S/N} = 10$~pixel$^{-1}$ at 4500~\AA, $\rm{S/N} =
25$~pixel$^{-1}$ at 5500~\AA, and $\rm{S/N} = 45$~pixel$^{-1}$ at
6500~\AA, comparable to that obtained for similarly faint stars by
\citet{koch08} and \citet*{koch09}.  We measure a velocity of $130.9
\pm 1.1$~\kms, consistent with the previous measurement of $132.7 \pm
2.2$~\kms\ from SG07, which suggests that Leo~IV-S1 does not have a
binary companion in a close enough orbit to affect its evolution or
abundances.

\section{ABUNDANCE ANALYSIS}
\label{results}

\subsection{Line Measurements and Atmospheric Parameters}

\begin{figure*}[t!]
\epsscale{1.20}
\plotone{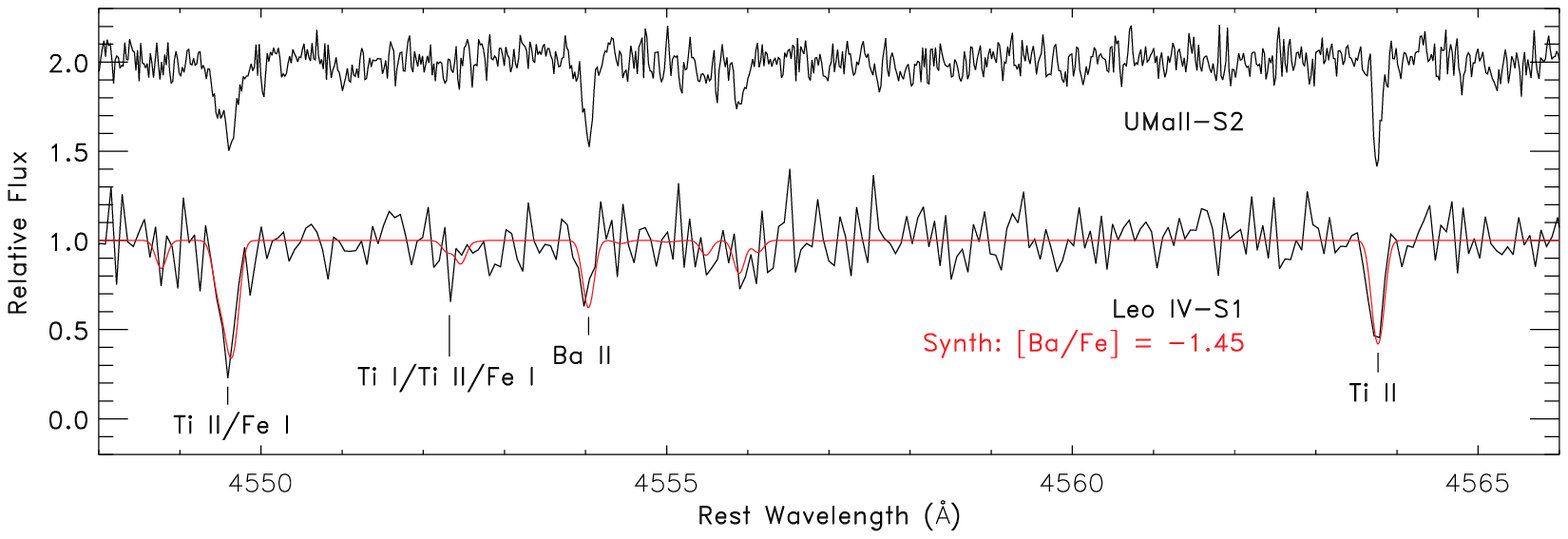}
\caption{Spectrum of Leo~IV-S1 around the
  \ion{Ba}{2}~$\lambda4554$~\AA\ line.  Despite the low S/N at these
  wavelengths, Ba is clearly detected.  Lines of Ti and Fe are also
  marked.  The red line is a synthetic spectrum from MOOG using the
  derived atmospheric parameters and abundance ratios, and the
  spectrum of UMaII-S2 ($T_{eff} = 4600$~K, $\rm{[Fe/H]} = -3.23$)
  from \citet{frebel09} is shown above for comparison.}
\label{specfig}
\end{figure*}

Using a line list taken from \citet{mcw95} and \citet{frebel09}, we
measured equivalent widths (EWs) with the IRAF task {\sc splot}.  We
detected $\sim50$ \ion{Fe}{1} lines, and between one and ten lines for
the following species: \ion{Fe}{2}, \ion{Mg}{1}, \ion{Ca}{1},
\ion{Sc}{2}, \ion{Ti}{1}, \ion{Ti}{2}, \ion {Na}{1}, \ion{Cr}{1},
\ion{Ni}{1}, \ion{Sr}{2}, and \ion{Ba}{2}.  A portion of the spectrum
illustrating the detection of Ba is displayed in Figure~\ref{specfig},
and the EWs of all measured lines are listed in Table~1.

Based on a combined photometric and spectroscopic analysis,
\citet{kirby08} estimated $T_{eff} = 4330$~K, $\log{g} = 1.0$, $\xi =
1.6$~\kms, and $\rm{[Fe/H]} = -2.9$ for Leo~IV-S1.  Starting with
these parameters, we constructed a 1D plane-parallel Kurucz model
atmosphere \citep{kurucz92} and then iteratively redetermined the
stellar parameters using the \ion{Fe}{1} lines with the 2009 version
of MOOG \citep{sneden73}.  We first established the microturbulent
velocity by minimizing the trend of \ion{Fe}{1} abundance with EW.
The derived value was $\xi = 3.2$~\kms, which would be quite high for
the less luminous stars that are frequently observed in the Milky Way
halo and globular clusters, but is comparable to values obtained for
cool EMP giants from low S/N spectra\footnote{Even if the S/N results
  in $\xi$ being overestimated, the effect on the abundances is small
  ($\sim0.1$~dex).} in a number of other studies
\citep{mcw95,koch08,aoki09,fks09}.

Next, we determined the effective temperature.  Leo~IV-S1 is
unfortunately too faint to have been detected by 2MASS, so the reddest
available color is $V-I$ (converted from the SDSS magnitudes using the
\citealt{jordi06} transformations).  The color-temperature relation of
\citet{alonso99} predicts $T_{eff} = 4330$~K using either $B-V$ (from
\citealt{moretti09}) or $V-I$.  At this temperature, there is still a
weak negative trend of \ion{Fe}{1} abundance with excitation
potential, as noted for similar stars by \citet[][and references
  therein]{norris09}, perhaps indicating deviations from local
thermodynamic equilibrium.  Forcing \ion{Fe}{1} excitation balance and
deriving $T_{eff}$ from the spectrum alone would lead to a lower value
($\sim$4200~K).

Ideally, the surface gravity would be set by imposing ionization
balance on the \ion{Fe}{1} and \ion{Fe}{2} lines.  Unfortunately, with
the relatively low S/N and resolution of our spectra, very few
\ion{Fe}{2} lines were detectable, and they are all either weak
features and/or in low S/N regions of the spectrum.  We therefore do
not consider any of our \ion{Fe}{2} measurements (which span nearly an
order of magnitude in abundance) to be very reliable.  We instead
resorted to the more basic technique of applying the Stefan-Boltzmann
law and Newton's law of gravitation to calculate the gravity.  The
apparent $r$ magnitude of Leo~IV-S1 after correcting for interstellar
reddening of $E(B-V) = 0.025$~mag \citep*{sfd98} is $r = 18.76$.
Given a distance modulus for Leo~IV of 20.94~mag \citep{moretti09},
the absolute magnitude is $M_{r} = -2.18$.  Using isochrones from
\citet{girardi04}, we estimate a bolometric correction of $-0.11$~mag
for stars of similar evolutionary state, yielding a luminosity of
637~L$_{\odot}$.  For a mass of 0.8~M$_{\odot}$ the corresponding
surface gravity is $\log{g} = 1.0$, varying only weakly with the
assumed temperature ($\Delta \log{g} = -0.05$~dex for $\Delta T_{eff}
= 100$~K).  This value for the gravity produces an \ion{Fe}{2}
abundance that is $\sim0.3$~dex higher than the \ion{Fe}{1} abundance,
but again, we do not regard the \ion{Fe}{2} measurement as reliable.
A much lower gravity of $\log{g} \sim 0$ would be required to bring
[\ion{Fe}{1}/H] and [\ion{Fe}{2}/H] into better agreement.  Our final
atmospheric parameters are therefore $T_{eff} = 4330$~K, $\log{g} =
1.0$, and $\xi = 3.2$~\kms, but we also derive abundances for the
purely spectroscopic values of $T_{eff} = 4200$~K, $\log{g} = 0.0$,
and $\xi = 3.2$~\kms\ for comparison.

\subsection{Derived Abundances and Uncertainties}

We list the measured abundances from MOOG in Table~\ref{abundances}.
Because of the low quality of our \ion{Fe}{2} measurement, we adopt
$\rm{[Fe/H]} = \rm{[}$\ion{Fe}{1}/H] to calculate [X/Fe] values (for
ionized species as well as neutral ones).  Note that we use the new
\citet{asplund09} solar abundances (with $\log{\epsilon\rm{(Fe)}} =
7.50$).

Since the photometric and spectroscopic solutions for $T_{eff}$ and
$\log{g}$ are not entirely consistent, assessing the impact that our
choices for these parameters have on the derived abundances is
important.  An estimate of the systematic uncertainties can be
obtained from the abundance differences between the two sets of
stellar parameters.  To quantify these further, we also vary the
atmospheric parameters one at a time by approximately their
uncertainties and examine the resulting abundance changes.  The
parameter uncertainties are set by considering how large a change is
allowed by the \ion{Fe}{1} abundances for $\xi$ and $T_{eff}$, and
assigning reasonable uncertainty levels for the gravity and overall
metallicity: $\Delta T_{eff} = +150$~K, $\Delta$log~g $ = -0.5$~dex,
$\Delta \mbox{[M/H]} = +0.3$~dex, and $\Delta \xi = +0.3$~\kms.  The
systematic uncertainties we derive are listed in
Tables~\ref{abundances} and \ref{uncertainties}.  Over this range of
parameters, the maximum iron abundance we obtain is $\rm{[Fe/H]} =
-2.96$, so we can confidently conclude that Leo~IV-S1 is indeed an EMP
star.

\section{DISCUSSION}
\label{discussion}

\subsection{The Frequency of EMP Stars in Dwarf Galaxies}

Including Leo~IV-S1, there are now detailed abundance studies (with
individually determined atmospheric parameters) for ten stars in the
ultra-faint dwarfs \citep{koch08,frebel09,norris09}.  The
\emph{highest} metallicity star included in these studies has
$\rm{[Fe/H]} = -2.0$ \citep{koch08}, and four have metallicities below
$\rm{[Fe/H]} = -3.0$.  As noted by \citet{frebel09}, with the
exception of Boo-1137 from \citet{norris09} these stars have been
selected independent of their metallicities: the sole selection
criterion (by necessity) is their apparent magnitude.  The 33\%
success rate (3 out of 9, after excluding Boo-1137) at finding EMP
stars strongly suggests that a substantial fraction of the stars in
these systems have extremely low metallicities \citep{kirby08,sf09}.
In order to obtain 3 EMP stars in a random drawing out of a sample of
9, the EMP fraction must be at least 10\% at the 95\% confidence
level.  The results of \citet{norris08} that 4 of 16 stars observed at
medium resolution in Boo~I (including Boo-1137) have metallicities
below $\rm{[Fe/H]} = -3.0$ provides further support for this case.  As
the observed data sets increase further, it therefore seems likely
that many more EMP stars, and perhaps stars with even lower
metallicities, will be identified.  Provided that one is willing to
invest the telescope time to obtain high-resolution spectra of
18th-19th magnitude stars, the ultra-faint dwarfs may be the most
promising targets for increasing samples of stars with $\rm{[Fe/H]} <
-3.5$ and studying the fossil clues left behind by the first
generation of stars.

\subsection{Abundance Patterns in the Ultra-Faint Dwarfs}

\begin{figure*}[t!]
\epsscale{1.20}
\plotone{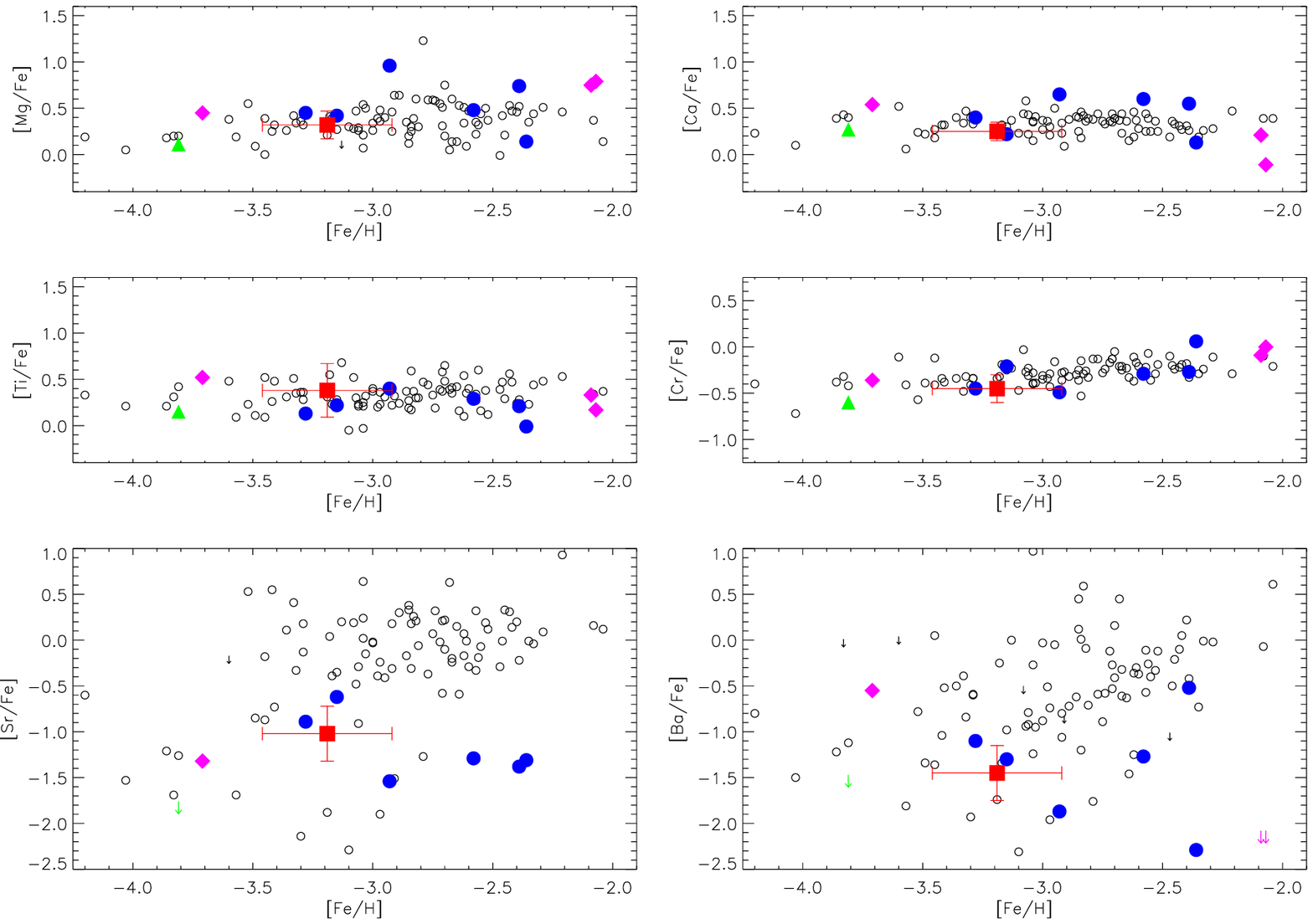}
\caption{Abundance pattern of Leo~IV-S1 (\emph{filled red square})
  compared to stars in other ultra-faint dwarf galaxies (\emph{filled
    blue circles}: ComBer and UMa~II from \citealt{frebel09};
  \emph{filled magenta diamonds}: Bo{\"o}tes~I and Hercules from
  \citealt{koch08} and \citealt{norris09}; \emph{filled green
    triangle}: Sculptor star S1020549 from \citealt{fks09}) and a
  representative sample of metal-poor Milky Way halo stars (\emph{open
    black circles}) from \citet{cayrel04}, \citet{cohen04},
  \citet{aoki05}, \citet{francois07}, and \citet{lai08}.  With the
  possible exception of the two relatively metal-rich stars in
  Hercules at $\mbox{[Fe/H]} = -2$, the distribution of $\alpha$ and
  iron-peak abundance ratios is very similar across all of the
  ultra-faint dwarfs and the halo.  All data displayed here have been
  adjusted to the \citet{asplund09} solar abundance scale.}
\label{patternfig}
\end{figure*}

The abundances of light and iron-peak elements in Leo~IV-S1 match
closely those that we measured in UMa~II and ComBer.  The $\alpha$
elements Mg, Ca, and Ti are each enhanced by $\sim0.3$~dex compared to
the solar ratios, identical within the uncertainties to those of the
two EMP stars in UMa~II.  Sc, Cr, and Ni in Leo~IV-S1 also agree with
the measured abundances of UMa~II below $\rm{[Fe/H]} = -3$.  Only a
conservative [C/Fe] limit could be obtained for Leo~IV-S1, indicating
that the star is not strongly C-enriched.  These similarities suggest
that whatever process is responsible for producing elements from Na at
least through the iron-peak in the ultra-faint dwarfs seems to be
nearly universal, yielding similar abundances in almost every star
examined so far.  The only exception is the ratio of hydrostatic to
explosive $\alpha$ elements (e.g., [Mg/Ca]), which is strongly
enhanced in a fraction of the ultra-faint dwarf stars
\citep{koch08,frebel09,feltzing09}.  As found by \citet{frebel09} and
illustrated in Fig.~\ref{patternfig}, this common abundance pattern in
ultra-faint dwarf stars (including Leo~IV-S1) also agrees well with
that of EMP stars in the Milky Way halo
\citep[e.g.,][]{cayrel04,lai08}.

Moreover, Leo~IV-S1 continues the trend of unusually low
neutron-capture abundances in the ultra-faint dwarfs
\citep{koch08,frebel09}, with $\rm{[Ba/Fe]} = -1.45$ and $\rm{[Sr/Fe]}
= -1.02$.  Unlike the lighter species, for heavy elements the
ultra-faints as a whole do not agree with typical halo behavior; the
halo has a higher mean abundance and spans a larger range of [nc/Fe]
at similar metallicities \citep[Fig.~\ref{patternfig}; also
  see][]{francois07,lai08}.  Stars in the brighter dwarf spheroidals
(dSphs) generally have roughly solar abundances of Ba and Eu, although
a few of the most metal-poor stars in those galaxies show a similar
deficiency of heavy elements as the ultra-faint dwarfs
\citep*{fulbright04,fks09}.  This distinction from both the halo and
the classical dSphs suggests that the heavy elements may be produced
differently in the ultra-faint dwarfs than in their brighter
counterparts (at least at later times).

Abundance measurements of the few strongly r-process enhanced EMP
stars in the halo indicate that SNe that produce r-process elements in
large quantities must be rare \citep[e.g.,][]{mcw95b} or inefficient
at dispersing those elements into the surrounding gas.  It has been
suggested that core-collapse SNe over a narrow mass range are the
astrophysical site for the main r-process, perhaps in the lowest mass
SNe ($8-10$~M$_{\odot}$) \citep[e.g.,][]{qw03,wanajo03}.  If the
enrichment of all of the ultra-faint dwarfs is a result of randomly
sampling supernovae from a common initial mass function (IMF), and
assuming the main r-process to be the dominant source for the observed
neutron-capture elements, then most dwarfs that incompletely sample
the SN mass function will show deficient [Sr/Fe] and [Ba/Fe] ratios
because r-process SNe are rare.  A small fraction, however, should
contain relatively r-process rich EMP stars.  Since the IMF of the
first stars is expected to be top-heavy \citep[e.g.,][]{yoshida06},
the low r-process abundances we observe could arise naturally if
r-process elements are predominantly made by these lower-mass SNe.

Despite the overall broad similarities with the other ultra-faints,
possible signs of stochasticity in the abundance patterns of the
faintest dwarfs are also evident in the data that have been acquired
over the past few years.  Two stars in Hercules and one each in Boo~I
\citep{feltzing09} and Draco \citep{fulbright04} exhibit very high
$\rm{[Mg/Ca]}$ ratios that are argued to result from small numbers of
supernovae and the resulting incomplete sampling of the IMF
\citep{koch08}.  The Hercules stars also have extremely low (nearly
unprecedented; see Fig. \ref{patternfig}) upper limits for [Ba/Fe],
while Ba has been detected in every star observed so far in UMa~II,
ComBer, and Leo~IV, despite much lower overall metallicities.  It may
be noteworthy that it is the \emph{most luminous} ultra-faint dwarfs
that seem to contain these unusual signatures, but larger samples in
all of these galaxies are needed before drawing strong conclusions.

\subsection{Supernovae and Nucleosynthesis in Leo~IV}

\begin{figure}[t!]
\epsscale{1.24}
\plotone{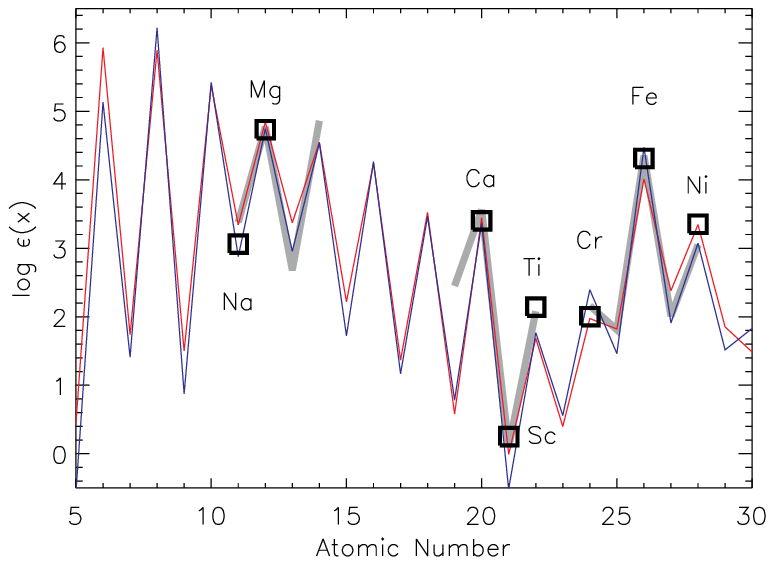}
\caption{Observed abundance pattern of Leo~IV-S1 \emph{(black
    squares)} compared to the \citet{cayrel04} halo sample
  \emph{(thick gray line)} at similar metallicities and Population III
  supernova nucleosynthesis models from \citet{hw08}.  A
  \citeauthor{hw08} ``standard'' 10.2~M$_{\odot}$ model (\emph{red
    line}), diluted by $4700$~M$_{\odot}$ of primordial gas,
  reproduces the overall abundance pattern fairly well.  A high
  explosion energy 29.5~M$_{\odot}$ \citeauthor{hw08} SN (added to
  $2.6 \times 10^{5}$~M$_{\odot}$ of hydrogen) predicts a similar
  pattern for many elements but with discrepant abundances for Sc and
  Cr (\emph{blue line}).}
\label{nucleosynthfig}
\end{figure}

\citet{hw08} have shown that Population~III SNe from initially
metal-free massive stars can produce an elemental abundance pattern
similar to that measured by \citet{cayrel04} for EMP halo stars.  In a
similar study, \citet*{tominaga07} concluded that Pop~III hypernovae
with unusually high energies are necessary to match the
\citet{cayrel04} data.  The agreement between the abundances of
Leo~IV-S1 and the \citeauthor{cayrel04} sample suggests that some form
of Pop~III nucleosynthesis may be able to explain the chemical
abundances of Leo~IV as well.  In Fig.~\ref{nucleosynthfig} we
demonstrate the quality of the match that can be obtained between the
observed abundances and the models; the best fit found by comparisons
with the grid of models from \citet{hw08} is for a low-mass
($\sim10$~M$_{\odot}$) SN with an average explosion energy.  Higher
mass hypernova explosions can also provide acceptable fits.

Because the number of stars in Leo~IV is so small ($L_{V} =
14000$~L$_{\odot}$), the metal content of the entire galaxy is
extremely low.  For example, given the mean metallicity determined by
\citet{kirby08} and assuming a stellar mass-to-light ratio of
1~M$_{\odot}$/L$_{\odot}$, Leo~IV contains just 0.042~M$_{\odot}$ of
Fe.  Since this is comparable to the amount of Fe produced by the
best-fitting Pop III SN models \citep{hw08}, if Leo~IV was indeed
enriched by such explosions then a \emph{single supernova may be
  enough to have produced all of the observed heavy elements}.  It is
also possible, of course, that multiple supernovae contributed to the
chemical evolution of the galaxy if most of the metals were blown out
via winds rather than having been incorporated into subsequent
generations of stars.  Nevertheless, we tentatively conclude that
Leo~IV-S1 may reflect the nucleosynthetic yields of the first Pop~III
star that the galaxy formed, at a time when its gas content was
$\sim4\times10^{4}$~M$_{\odot}$.  Other ultra-faint dwarfs therefore
might reveal the abundance patterns of SNe with different masses,
consistent with recent observations of Hercules \citep{koch08}

\section{SUMMARY AND CONCLUSIONS}
\label{conclusions}

We have carried out a high-resolution abundance analysis for the
brightest star in the ultra-faint dwarf galaxy Leo~IV.  With
$\rm{[Fe/H]} = -3.2$, Leo~IV-S1 adds to the rapidly increasing sample
of extremely metal-poor stars in dwarf galaxies.  The low metallicity
of Leo~IV-S1 provides further support for the hypothesis that a
substantial fraction ($\gtrsim 10$\%) of the stars in the faintest
dwarfs lie in the EMP regime.

The abundance pattern in Leo~IV is extremely similar to that found in
both the other ultra-faint dwarfs and the metal-poor Milky Way halo.
As suggested by \citet{frebel09}, this excellent agreement
demonstrates that the metal-poor end of the halo metallicity
distribution could have been formed in galaxies like the ultra-faint
dwarfs.  The only exception to the close match between the halo and
the ultra-faint dwarfs is the neutron-capture elements, which still
appear somewhat lower in the dwarfs, but more measurements of these
elements are needed.

Interestingly, the most metal-poor stars in some of the brighter dSphs
seem to share the same chemical signature of $\alpha$-enhancement and
neutron-capture depletion \citep{fulbright04,fks09}, although the
abundances in those systems deviate substantially at higher
metallicities \citep{shetrone03,venn04}.  The similarity between the
abundances of Leo~IV-S1 and other dwarfs such as UMa~II, ComBer,
Boo~I, and Sculptor suggests that the initial enrichment in many
galaxies may have been universal.  Differences between the abundances
in the faintest dwarfs (ComBer, UMa~II, and Leo~IV) and the stars in
somewhat more luminous systems, on the other hand, could point to
stochastic chemical variations.  Finally, we show that the abundance
pattern of Leo~IV-S1 is consistent with Population III supernova
models, raising the possibility that Leo~IV was enriched by some of
the first stars.

\acknowledgements{We thank Dan Kelson for assistance with the data
  reduction, Steve Shectman for observing suggestions, Alexander Heger
  for extensive help with comparisons to theoretical SN models, and
  the referee for a helpful report.  JDS gratefully acknowledges the
  support of a Vera Rubin Fellowship provided by the Carnegie
  Institution of Washington, and AF that of a Clay Fellowship
  administered by the Smithsonian Astrophysical Observatory.  This
  work was also supported by NASA through Hubble Fellowship grant
  HST-HF-01233.01 awarded to ENK by the Space Telescope Science
  Institute, which is operated by the Association of Universities for
  Research in Astronomy, Inc., for NASA, under contract NAS 5-26555.
  This research made use of NASA's Astrophysics Data System
  Bibliographic Services.}

{\it Facilities:} \facility{Magellan:Clay (MIKE)}

\LongTables
\begin{deluxetable}{lllrrr}
\tablecolumns{6}
\tablewidth{0pt}
\tabletypesize{\scriptsize}
\tablecaption{\label{Tab:Eqw} Equivalent width measurements}
\tablehead{
\colhead{Element} & \colhead{$\lambda$} &\colhead{$\chi$} &\colhead{$\log{gf}$}&
\colhead{EW}&\colhead{$\log{\epsilon}$}\\
\colhead{} & \colhead{[\AA]} &\colhead{[eV]} &\colhead{}&
\colhead{[m\AA]}&\colhead{}
}
\startdata
Na\,I  &  5889.973 &  0.00 &  \phs0.100 &  220.0 & \phs3.28 \\
Na\,I  &  5895.940 &  0.00 &  $-$0.200  &  160.0 & \phs2.84 \\
Mg\,I  &  4571.102 &  0.00 &  $-$5.670  &   64.0 & \phs4.47 \\
Mg\,I  &  4703.003 &  4.34 &  $-$0.520  &   75.0 & \phs5.11 \\
Mg\,I  &  5172.698 &  2.71 &  $-$0.380  &  231.0 & \phs4.66 \\
Mg\,I  &  5183.619 &  2.72 &  $-$0.160  &  256.0 & \phs4.66 \\
Mg\,I  &  5528.418 &  4.34 &  $-$0.341  &   74.0 & \phs4.76 \\
Ca\,I  &  5588.764 &  2.52 &  \phs0.358 &   47.0 & \phs3.39 \\
Ca\,I  &  5594.471 &  2.52 &  \phs0.097 &   22.0 & \phs3.22 \\
Ca\,I  &  6102.727 &  1.88 &  $-$0.770  &   38.0 & \phs3.51 \\
Ca\,I  &  6122.226 &  1.89 &  $-$0.320  &   62.0 & \phs3.38 \\
Ca\,I  &  6162.180 &  1.90 &  $-$0.090  &   90.0 & \phs3.47 \\
Ca\,I  &  6439.083 &  2.52 &  \phs0.390 &   61.0 & \phs3.45 \\
Sc\,II &  5031.024 &  1.36 &  $-$0.400  &   41.0 & \phs0.27 \\
Sc\,II &  5526.821 &  1.77 &  \phs0.020 &   42.0 & \phs0.33 \\
Sc\,II &  5657.880 &  1.51 &  $-$0.600  &   18.0 & \phs0.14 \\
Ti\,I  &  5039.964 &  0.02 &  $-$1.130  &   25.0 & \phs1.76 \\
Ti\,I  &  5210.392 &  0.05 &  $-$0.884  &   58.0 & \phs2.01 \\
Ti\,II &  4417.723 &  1.16 &  $-$1.430  &   82.0 & \phs2.13 \\
Ti\,II &  4443.812 &  1.08 &  $-$0.700  &  127.0 & \phs1.95 \\
Ti\,II &  4468.500 &  1.13 &  $-$0.600  &  120.0 & \phs1.79 \\
Ti\,II &  4501.278 &  1.12 &  $-$0.750  &  136.0 & \phs2.15 \\
Ti\,II &  4563.766 &  1.22 &  $-$0.960  &  122.0 & \phs2.24 \\
Ti\,II &  4571.982 &  1.57 &  $-$0.530  &  111.0 & \phs2.12 \\
Ti\,II &  4589.953 &  1.24 &  $-$1.790  &   70.0 & \phs2.37 \\
Ti\,II &  5185.908 &  1.89 &  $-$1.350  &   26.0 & \phs2.02 \\
Ti\,II &  5336.794 &  1.58 &  $-$1.700  &   63.0 & \phs2.48 \\
Ti\,II &  5381.028 &  1.57 &  $-$2.080  &   21.0 & \phs2.19 \\
Cr\,I  &  4254.346 &  0.00 &  $-$0.114  &  141.0 & \phs1.92 \\
Cr\,I  &  5345.807 &  1.00 &  $-$0.980  &   26.0 & \phs2.08 \\
Fe\,I  &  4447.728 &  2.22 &  $-$1.339  &   77.0 & \phs4.30 \\
Fe\,I  &  4494.573 &  2.20 &  $-$1.136  &   74.0 & \phs4.01 \\
Fe\,I  &  4871.325 &  2.86 &  $-$0.362  &   60.0 & \phs3.81 \\
Fe\,I  &  4872.144 &  2.88 &  $-$0.567  &   91.0 & \phs4.44 \\
Fe\,I  &  4890.763 &  2.87 &  $-$0.394  &   71.0 & \phs3.99 \\
Fe\,I  &  4891.502 &  2.85 &  $-$0.111  &   83.0 & \phs3.83 \\
Fe\,I  &  4994.138 &  0.91 &  $-$2.956  &   71.0 & \phs3.93 \\
Fe\,I  &  5041.076 &  0.96 &  $-$3.086  &   86.0 & \phs4.29 \\
Fe\,I  &  5041.763 &  1.48 &  $-$2.203  &  114.0 & \phs4.46 \\
Fe\,I  &  5049.827 &  2.28 &  $-$1.355  &   67.0 & \phs4.09 \\
Fe\,I  &  5123.730 &  1.01 &  $-$3.058  &  115.0 & \phs4.66 \\
Fe\,I  &  5127.368 &  0.91 &  $-$3.249  &   81.0 & \phs4.31 \\
Fe\,I  &  5133.699 &  4.18 &  \phs0.140 &   23.0 & \phs4.38 \\
Fe\,I  &  5150.852 &  0.99 &  $-$3.037  &  113.0 & \phs4.58 \\
Fe\,I  &  5151.917 &  1.01 &  $-$3.321  &   70.0 & \phs4.38 \\
Fe\,I  &  5166.284 &  0.00 &  $-$4.123  &   94.0 & \phs4.08 \\
Fe\,I  &  5171.610 &  1.48 &  $-$1.721  &  118.0 & \phs3.98 \\
Fe\,I  &  5191.465 &  3.04 &  $-$0.551  &   47.0 & \phs4.01 \\
Fe\,I  &  5192.353 &  3.00 &  $-$0.421  &   83.0 & \phs4.27 \\
Fe\,I  &  5194.949 &  1.56 &  $-$2.021  &  120.0 & \phs4.41 \\
Fe\,I  &  5216.283 &  1.61 &  $-$2.082  &   91.0 & \phs4.17 \\
Fe\,I  &  5225.534 &  0.11 &  $-$4.755  &   82.0 & \phs4.72 \\
Fe\,I  &  5232.952 &  2.94 &  $-$0.057  &  109.0 & \phs4.13 \\
Fe\,I  &  5250.216 &  0.12 &  $-$4.938  &   68.0 & \phs4.76 \\
Fe\,I  &  5281.798 &  3.04 &  $-$0.833  &   48.0 & \phs4.29 \\
Fe\,I  &  5283.629 &  3.24 &  $-$0.524  &   40.0 & \phs4.13 \\
Fe\,I  &  5302.307 &  3.28 &  $-$0.720  &   55.0 & \phs4.57 \\
Fe\,I  &  5307.369 &  1.61 &  $-$2.912  &   48.0 & \phs4.48 \\
Fe\,I  &  5324.191 &  3.21 &  $-$0.103  &   99.0 & \phs4.39 \\
Fe\,I  &  5497.526 &  1.01 &  $-$2.825  &  128.0 & \phs4.49 \\
Fe\,I  &  5501.477 &  0.96 &  $-$3.046  &  110.0 & \phs4.41 \\
Fe\,I  &  5506.791 &  0.99 &  $-$2.789  &  142.0 & \phs4.60 \\
Fe\,I  &  5569.631 &  3.42 &  $-$0.500  &   54.0 & \phs4.49 \\
Fe\,I  &  5572.851 &  3.40 &  $-$0.275  &   57.0 & \phs4.27 \\
Fe\,I  &  5615.658 &  3.33 &  \phs0.050 &   82.0 & \phs4.14 \\
Fe\,I  &  6136.624 &  2.45 &  $-$1.410  &   82.0 & \phs4.37 \\
Fe\,I  &  6137.702 &  2.59 &  $-$1.346  &   66.0 & \phs4.32 \\
Fe\,I  &  6191.571 &  2.43 &  $-$1.416  &   95.0 & \phs4.48 \\
Fe\,I  &  6219.287 &  2.20 &  $-$2.448  &   40.0 & \phs4.58 \\
Fe\,I  &  6230.736 &  2.56 &  $-$1.276  &   60.0 & \phs4.13 \\
Fe\,I  &  6252.565 &  2.40 &  $-$1.767  &   59.0 & \phs4.39 \\
Fe\,I  &  6265.141 &  2.18 &  $-$2.550  &   21.0 & \phs4.31 \\
Fe\,I  &  6335.337 &  2.20 &  $-$2.180  &   43.0 & \phs4.34 \\
Fe\,I  &  6393.612 &  2.43 &  $-$1.576  &   68.0 & \phs4.33 \\
Fe\,I  &  6400.009 &  3.60 &  $-$0.290  &   40.0 & \phs4.24 \\
Fe\,I  &  6411.658 &  3.65 &  $-$0.595  &   25.0 & \phs4.35 \\
Fe\,I  &  6421.360 &  2.28 &  $-$2.014  &   44.0 & \phs4.28 \\
Fe\,I  &  6430.856 &  2.18 &  $-$1.946  &   75.0 & \phs4.44 \\
Fe\,I  &  6494.994 &  2.40 &  $-$1.239  &  100.0 & \phs4.28 \\
Fe\,I  &  6677.997 &  2.69 &  $-$1.420  &   58.0 & \phs4.38 \\
Fe\,I  &  6750.164 &  2.42 &  $-$2.621  &   13.0 & \phs4.43 \\
Fe\,II &  4923.930 &  2.89 &  $-$1.260  &   84.0 & \phs4.02 \\
Fe\,II &  5018.450 &  2.89 &  $-$1.110  &  120.0 & \phs4.35 \\
Fe\,II &  5197.560 &  3.23 &  $-$2.220  &   44.0 & \phs4.83 \\
Fe\,II &  5276.000 &  3.20 &  $-$2.010  &   67.0 & \phs4.88 \\
Fe\,II &  6456.391 &  3.90 &  $-$2.075  &   21.0 & \phs5.00 \\
Ni\,I  &  6643.638 &  1.68 &  $-$2.300  &   48.0 & \phs3.56 \\
Ni\,I  &  6767.784 &  1.83 &  $-$2.170  &   20.0 & \phs3.15 \\
Sr\,II &  4215.539 &  0.00 &  $-$0.170  &  120.0 & $-$1.34 \\
Ba\,II &  4554.036 &  0.00 &  \phs0.160 &   62.0 & $-$2.46
\enddata
\label{ew_table}
\end{deluxetable}

\begin{deluxetable}{lclcccc}
\tablewidth{0pt}
\tabletypesize{\scriptsize}
\tablecolumns{7}
\tablecaption{Stellar Parameters and Abundances for Leo~IV-S1}
\tablehead{
  \multicolumn{1}{c}{} & \multicolumn{5}{c}{Fiducial Model}  & 
  \multicolumn{1}{c}{Spectroscopic Model} \\
  \multicolumn{1}{c}{} & 
  \multicolumn{5}{c}{$\overline{\phm{SpanningSpanningSpanningSpanningSpanningSpanni}}$} &
  \multicolumn{1}{c}{$\overline{\phm{SpanningSpanningg}}$} \\
\colhead{Parameter/} & \colhead{Value/} & 
\colhead{$\log{\epsilon(X)}$} &
\colhead{N$_{\mbox{lines}}$} & 
\colhead{$\sigma_{\mbox{stat}}$\tablenotemark{a}} &
\colhead{$\sigma_{\mbox{sys}}$\tablenotemark{b}} & 
\colhead{Value/} \\
\colhead{Species} & \colhead{Abundance Ratio} & 
\colhead{} & \colhead{} & \colhead{} & 
\colhead{} & \colhead{Abundance Ratio}
}
\startdata 
$M_{r}$            & $-2.18$\phd   &     &    &       &   & $-2.18$\phd \\  
$T_{eff}$ [K]      & \phn4330\phd  &     &    &       &   & \phn4200\phd  \\
$\log{g}$         & \phn\phn1.0\phn &   &    &       &   & \phn\phn0.0\phn     \\
$\xi$ [\kms]      & \phn\phn3.2\phn &   &    &       &   & \phn\phn3.2\phn \\
$\rm{[Fe~I/H]}$   & $-3.19$\phd  & \phs4.31 & 50 & 0.03  & 0.27  & $-3.09$\phd  \\
$\rm{[Fe~II/H]}$  & $-2.88$:  & \phs4.62: & 5  & 0.19  & 0.14  & $-2.89$:  \\
$\rm{[C/Fe]}$    & $<$$-0.08$\phs\phd & $<$~5.16 & synth  & ...  & ...  &  $<0.32$\phn\phd \\ 
$\rm{[Na~I/Fe]}$  & \phs0.01\phd & \phs3.06 & 2  & 0.28  & 0.09  & $-0.04$\phd  \\
$\rm{[Mg~I/Fe]}$  & \phs0.32\phd & \phs4.73 & 5  & 0.12  & 0.09  & \phs0.62\phd \\
$\rm{[Ca~I/Fe]}$  & \phs0.25\phd & \phs3.40 & 6  & 0.05  & 0.09  & \phs0.26\phd \\
$\rm{[Ti~I/Fe]}$  & \phs0.13\phd & \phs1.89 & 2  & 0.16  & 0.07  & \phs0.08\phd \\
$\rm{[Ti~II/Fe]}$ & \phs0.38\phd & \phs2.14 & 10 & 0.06  & 0.28  & \phs0.40\phd \\
$\rm{[Sc~II/Fe]}$ & \phs0.29\phd & \phs0.25 & 3  & 0.06  & 0.33  & \phs0.00\phd \\
$\rm{[Cr~I/Fe]}$  & $-0.45$\phd  & \phs2.00 & 2  & 0.10  & 0.11  & $-0.44$\phd \\
$\rm{[Ni~I/Fe]}$  & \phs0.32\phd & \phs3.35 & 2  & 0.26  & 0.07  & \phs0.16\phd\\
$\rm{[Sr~II/Fe]}$ & $-1.02$\phd  & $-1.34$ & 1  & ...   & 0.26  & $-0.96$\phd \\
$\rm{[Ba~II/Fe]}$ & $-1.45$\phd  & $-2.46$ & 1  & ...   & 0.24  & $-1.56$\phd \\

\enddata
\label{abundances}

\tablecomments{All abundance ratios [X/Fe] (including ionized species)
  are calculated relative to Fe~I.}
\tablenotetext{a}{Statistical uncertainties are defined as the
  standard error of the mean of abundances of individual lines
  (accounting for small sample sizes where fewer than 10 lines are
  used).}
\tablenotetext{b}{Systematic uncertainties refer to quadrature sums of
  the changes listed in Table \ref{uncertainties} relative to
  \ion{Fe}{1} for each species.}
\end{deluxetable}

\begin{deluxetable}{lccccc}
\tablewidth{0pt}
\tabletypesize{\scriptsize}
\tablecolumns{6}
\tablecaption{Abundance Uncertainties}
\tablehead{
\colhead{Species} & \colhead{$\Delta \log{\epsilon(X)}$} & 
\colhead{$\Delta \log{\epsilon(X)}$} & 
\colhead{$\Delta \log{\epsilon(X)}$} & 
\colhead{$\Delta \log{\epsilon(X)}$} & \colhead{} \\
\colhead{} & \colhead{for $T_{eff} + 150$ K} & 
\colhead{for $\log{g} - 0.5$ dex} & 
\colhead{for $\mbox{[M/H]} + 0.3$} & 
\colhead{for $\xi + 0.3$ \kms} & \colhead{} 
}
\startdata 
Fe~I   & \phs0.23 & \phs0.13 & $-0.03$  & $-0.04$ \\
Fe~II  & $-0.08$  & $-0.10$  & \phs0.01 & $-0.05$ \\
Na~I   & \phs0.24 & \phs0.14 & $-0.07$  & $-0.12$ \\
Mg~I   & \phs0.19 & \phs0.21 & $-0.05$  & $-0.06$ \\
Ca~I   & \phs0.15 & \phs0.10 & $-0.02$  & $-0.01$ \\
Ti~I   & \phs0.29 & \phs0.14 & $-0.04$  & $-0.01$ \\
Ti~II  & $-0.01$  & \phs0.00 & \phs0.01 & $-0.06$ \\
Sc~II  & \phs0.02 & $-0.12$  & \phs0.02 & $-0.01$ \\
Cr~I   & \phs0.25 & \phs0.21 & $-0.09$  & $-0.07$ \\
Ni~I   & \phs0.25 & \phs0.07 & $-0.01$  & $-0.01$ \\
Sr~II  & $-0.01$ & \phs0.07 & $-0.04$  & $-0.11$ \\
Ba~II  & \phs0.07 & $-0.04$  & \phs0.02 & $-0.02$ \\

\enddata
\label{uncertainties}
\end{deluxetable}

\end{document}